\documentclass[aps,prb,preprint,superscriptaddress,showpacs]{revtex4}
\usepackage{graphicx}
\usepackage{dcolumn}
\usepackage{subfigure}
\usepackage[usenames]{color}

\begin{document}


\title{Long-lived submicrometric bubbles in very diluted alkali halide water solutions}

\author{Eug\`ene Duval \footnote{Corresponding author, eugene.duval@univ-lyon1.fr}}
\author{Sergey Adichtchev \footnote{On leave from the IA\&E Russian Academy of Sciences, Novosibirsk 630090 Russia}}
\author{Sergey Sirotkin}
\author{Alain Mermet}
\affiliation{Laboratoire de Physico-Chimie des Mat\'eriaux Luminescents,
Universit\'{e} Lyon 1 \\UMR-CNRS 5620 10, rue Ada Byron 69622
Villeurbanne Cedex, France}

\date{\today}

\begin{abstract}
Solutions of LiCl and of NaCl in ultrapure water were studied through Rayleigh/Brillouin scattering as a function of the concentration (molarity, M) of dissolved salt from 0.2M to extremely low concentration (2.10$^{-17}$M ).  The Landau-Placzek ratio, $R/B$, of the Rayleigh scattering intensity over the total Brillouin, was measured thanks to the dynamically controlled stability of the used Fabry-Perot interferometer. It was observed that the $R/B$ ratio follows two stages as a function of increasing dilution rate: after a strong decrease between	0.2M and 2.10$^{-5}$M, it increases to reach a maximum between 10$^{-9}$M and 10$^{-16}$M. The first stage corresponds to the decrease of the Rayleigh scattering by the ion concentration fluctuations with the decrease of salt concentration. The second stage, at lower concentrations, is consistent with the increase of the Rayleigh scattering by long-lived sub-microscopic bubbles with the decrease of ion concentration. The origin of these sub-microscopic bubbles is  the shaking of the solutions which was carried out after each centesimal dilution. The very long lifetime of the sub-microscopic bubbles and the effects of aging originate in the electric charge of bubbles. The increase of $R/B$ with the decrease of the low salt concentration corresponds to the increase of the sub-microscopic bubble size with the decrease of concentration, that is imposed by the bubble stability
due  to the covering of the surface bubble by negative ions.
\end{abstract}

\maketitle

\section{Introduction}

Due to the small size of H$_{2}$O molecules and to the hydrogen-bonding, liquid water is a very complex system with very peculiar properties. On the other hand, the structure of liquid water is very sensitive to dissolved molecules or ions. 
The structural effects on water of dissolved ions have been extensively studied by neutron scattering  \cite{enderby,botti,bouazizi,mancinelli}. They have consequences on the physical properties of water. Many of these have been studied as a function of the concentration of dissolved elements \cite{weissenborn,craig} :
surface tension, viscosity, gas solubilities,  etc. However, the structural effects and properties were studied for relatively high concentrations of ions or of salt (larger than 10$^{-2}$M), probably because lower ion concentrations have a negligible effect on the hydrogen-bonding structure.
Even if a low concentration of a dissolved salt has a non-measurable effect on the hydrogen-bonding structure, it was shown that dissolved ions in very low concentration have a crucial effect on the stability of sub-microscopic bubbles in water \cite{bunkin97,bunkin2009}. The structure and the properties of water around the stable sub-micorscopic bubbles are modified: the air/water interface of the bubbles changes the local chemical and biochemical reactive properties which can induce chemical or biochemical processes.

From the above remarks, it appears interesting to study the stability of sub-microscopic bubbles as a function of very weakly concentrated dissolved salts. Solutions of lithium chloride and sodium chloride in pure water, from 0.2M to 2.10$^{-17}$M, were obtained by successive centesimal dilutions. The solutions were vigorously shaken after each centesimal dilution to produce microbubbles. The technique that was used to detect the presence of sub-microscopic bubbles is polarized Rayleigh (or Rayleigh-Gans) light scattering \cite{hulst}. By this technique, it is possible to measure the intensity of light that is elastically scattered by sub-microscopic heterogeneities. The concomitant recording of the light scattering due to acoustical modes, that is the Brillouin scattering, allows to normalize the Rayleigh scattering intensity by taking the Landau-Placzek ratio (noted as $R/B$) of the Rayleigh intensity to the total Brillouin intensity. Because the  Brillouin scattering does not depend on the $\it{low}$ concentration of heterogeneities, the comparison of the Landau-Placzek ratios from different dilutions gives a relative measure of their respective Rayleigh scattering intensities.

After the presentation of the experimental results focusing on the dependence of the Landau-Placzek ratio on the concentration of alkali halide, as well as the effect of aging, the observed Rayleigh scattering will be interpreted by the presence of elecrically charged sub-microscopic bubbles in the aqueous salt dilutions. It will be shown that the increase of the Rayleigh scattering with the decrease of the salt concentration in the high dilution regime is rooted in an increase of the bubble size with decreasing salt concentration, making them stable due to electric interactions.

\section{Rayleigh and Brillouin light scattering. Landau-Placzek ratio}

In pure liquids, light is elastically scattered (Rayleigh scattering) by dynamical fluctuations of density. It is well-known that in pure water, due to the approximate equality of the specific heats, respectively at constant pressure and at constant volume, the intensity of Rayleigh scattering from density ($\rho$) fluctuations, $I_{R}(\rho)$, is very weak at room temperature \cite{cummins}. 

In aqueous solutions, for instance of salt, light is also elastically scattered by fluctuations of ion concentration. The dependence $I_{R}(C)$ of the Rayleigh scattering intensity on the weight concentration $C$ is expressed as \cite{dubois}:

\begin{equation}
\label{eq.1}
I_{R}(C)\propto n^{2} \left(\frac {\partial n}{\partial C}\right)^{2}_{P,T}<\Delta C^{2}>_{P,T}
\end{equation}

\noindent where $n$ is the refractive index and $<\Delta C^{2}>_{P,T}$ the mean square value of the fluctuations of $C$ at constant pressure and temperature \cite{dubois}. For $C<<1$, $<\Delta C^{2}>_{P,T}$ is simply proportional to $C$. \cite{dubois} The dependence on the angles specifying the direction of scattering relatively to the incidence direction is not given in Eq.~\ref{eq.1}, and  in the following expressions too.
 
Submicroscopic heterogeneities in aqueous solutions, like solid particles or bubbles, elastically scatter light. The dependence of the Rayleigh scattering, $I_{R}(V)$, on the number $N$ of heterogeneities per unit volume and on the heterogeneity volume $V$ is expressed as follows \cite{hulst}:

\begin{equation}
\label{eq.2}
I_{R}(V)\propto n_{\circ}^{2}\Delta n^{2}NV^{2}
\end{equation}

\noindent where $n_{\circ}$ is the refractive index of the aqueous solution and $\Delta n$ is  the difference between the refractive indices of  heterogeneity and of the  solution. Eq.~\ref{eq.2} is valid in the case of non-coherence between light waves scattered by different heterogeneities. If there exists coherence, $N$ must be replaced by $N^{2}$. The coherence in the Rayleigh scattering is expected in the case of an ordered (for instance, periodic) arrangement of heterogeneities.

In Brillouin scattering, light is inelastically scattered by density fluctuations which are induced by sound waves propagating in the liquid. The Brillouin frequency  shift $\Delta \nu$ (shift of the Brillouin line relatively to the Rayleigh line) is proportional to the sound velocity, $v$. The expression of $\Delta \nu$ as a function of 
$v$ is \cite{cummins}:

\begin{equation}
\label{eq.3}
\Delta \nu=\frac{v}{\lambda_{\circ}}2n_{\circ}\sin{\theta /2} 
\end{equation}

\noindent where $\lambda_{\circ}$ is the vacuum wavelength of incident light and $\theta$ the angle between the scattering direction and the incidence one.

The dependence on the sound velocity  of the Brillouin intensity $I_{B}$, that is derived from the relation between the sound velocity and the adiabatic hypersonic compressibility, is as follows \cite{cummins}:

\begin{equation}
\label{eq.4}
I_{B}\propto n_{\circ}\left(\frac{\partial n}{\partial \rho}\right)_{T} \frac{1}{v^{2}}
\end{equation}

The Landau-Placzek ratio, $R/B$, is the total Rayleigh intensity, $I_{R}$, divided by the total intensity of Brillouin scattering (Stokes plus antiStokes):

\begin{equation}
\label{eq.5}
\frac{R}{B}=\frac{I_{R}}{2I_{B}}=\frac{I_{R}(\rho)+I_{R}(C)+I_{R}(V)}{2I_{B}}
\end{equation}

\noindent If the solution is changed without noticeable modification of its refractive index and of sound velocity, from Eq.~\ref{eq.4}, $I_{B}$ remains unchanged, so that the Brillouin scattering intensity allows to normalize the Rayleigh scattering intensity.

\section{Experimental}

\subsection{Samples} 

The water that was used for the dilutions was submitted to different treatments: it was demineralised, purified from organic matter, filtered with a porosity of 0.2 $\mu m$ and sterilised by ultraviolet radiation exposure. Its conductivity was 0.056 $\mu S/cm$  and its TOC (total organic carbon) was less than 10 ppb. Before the preparation of the solutions, the water was stored for about two days in bottles made of Schott-Duran glass that is known to minimize the release of glass species into the contained water. All the recipients (storing bottles, storing flasks, light measurement cuvettes), that were used in this study, were carefully cleaned and sterilized by heating at 160$^{\circ}$ C for two hours. 

The solutions of LiCl and NaCl were obtained by successive dilutions with a ratio of 100, i.e. by successive $\it{centesimal}$ dilutions. The initial solution was obtained by dissolving 1 g of salt in 100 cm$^{3}$ of ultrapure water, i.e., 10$^{-2}$ g cm$^{-3}$ ($\approx 0.2M$), downto 10$^{-18}$ g cm$^{-3}$ ($\approx 2.10^{-17}$M). The different diluted solutions are noted C1 for 10$^{-2}$g cm$^{-3}$ to C9 for 10$^{-18}$g cm$^{-3}$. The used salts were ultra-dry and ultrapure (99.995$\%$ for LiCl and 99.999$\%$ for NaCl).

The centesimal dilutions were operated upon precise shaking condtions: each dilution step (from C1 to C9) was terminated with a vigorous shaking of 150 strokes during 7.5 s. For the purpose of the study, the solutions were sumitted to a second identical shaking some days after the first dilution-shaking process.  

For each solution, reference samples made of only ultrapure water were prepared following the very \textit{same} conditions of dilution and shaking as the salt solutions (water "`diluted"' in water).

In order to study the effect of the glass container, series of diluted solutions were prepared in both borosilicate and Schott-Duran flasks, the latter being known to least release ions in the solution. The borosilicate flasks and the Schott-Duran ones had a volume of, respectively, 30 cm$^{3}$ and 67 cm$^{3}$. Both types of flasks, which were filled up to the 2/3 of their volume, had a square section. The shaking created a strong visible bubbling of the solution. The dilution-shaking procedure was carried out in purified air in Laboratoire Boiron under our supervision . The solutions were stored several months at 20$^{\circ}$C.

For the light scattering measurements, a small volume of a solution was taken from the glass flasks and injected with sterilized microsyringes in parallelepipedic cuvettes (10x10x46 mm$^{3}$) with optically-polished Suprasil walls. The solutions remaining in the glass flasks were kept in the dark  for 5 days or 12 days and then shaken a second time (using the same shaking time and rate as in the first shaking). 

\subsection{Technique}

The Rayleigh and Brillouin scattering experiments were performed using the 532 nm line of a continuous yttrium aluminium garnet (YAG) laser for the excitation of the sample and a six-pass tandem Fabry-Perot interferometer coupled with a Si avalanche photodiode for the detection of the scattered light. The laser power was not higher than 35 mW ($\approx$100 mW/mm$^{2}$). A twice larger input power did not change the $R/B$ ratio. This demonstrates that the used laser power did not induce cavitation. In order to suppress the possible spilling of photons elastically scattered by the cuvette walls, the scattering region was spatially filtered out using a 6x2 mm$^{2}$ opaque screen aperture. The scattered light was collected perpendicularly to the incident direction ($\theta=\pi /2$). The temperature of the laboratory was 20$^{\circ}$ C.

The $R/B$ spectra shown thereafter result from the accumulation of 300 identical scans (1 scan takes about 0.5 s), which ensures a good signal-to-noise ratio. As observed by eye, stochastic intense scattering (flashes), probably occuring from bubbles with sizes larger than a few microns (as determined  experimentally through different pore size filterings and observations with an optical microscope), was identified as a possible spurious contribution to the true $R/B$ scattering. In order to discard this contribution, a discriminating counting procedure was used such that scans with elastic scattering rates (within the range of $\pm$1~GHz) exceeding the Brillouin line average intensities were rejected in the accumulation. The number of rejected scans was recorded as a relevant parameter, through the ratio $F_{rej}$, defined as the number of spectra that were rejected by the discriminator divided by the total number of accumulated spectra. Typical values of the rejection rate did not exceed $10\%$.
	
 \subsection{Experimental results}

\subsubsection{Dependence of Landau-Placzek ratio on the very low alkali halide concentration}
 
\begin{figure}
\includegraphics[width=\columnwidth]{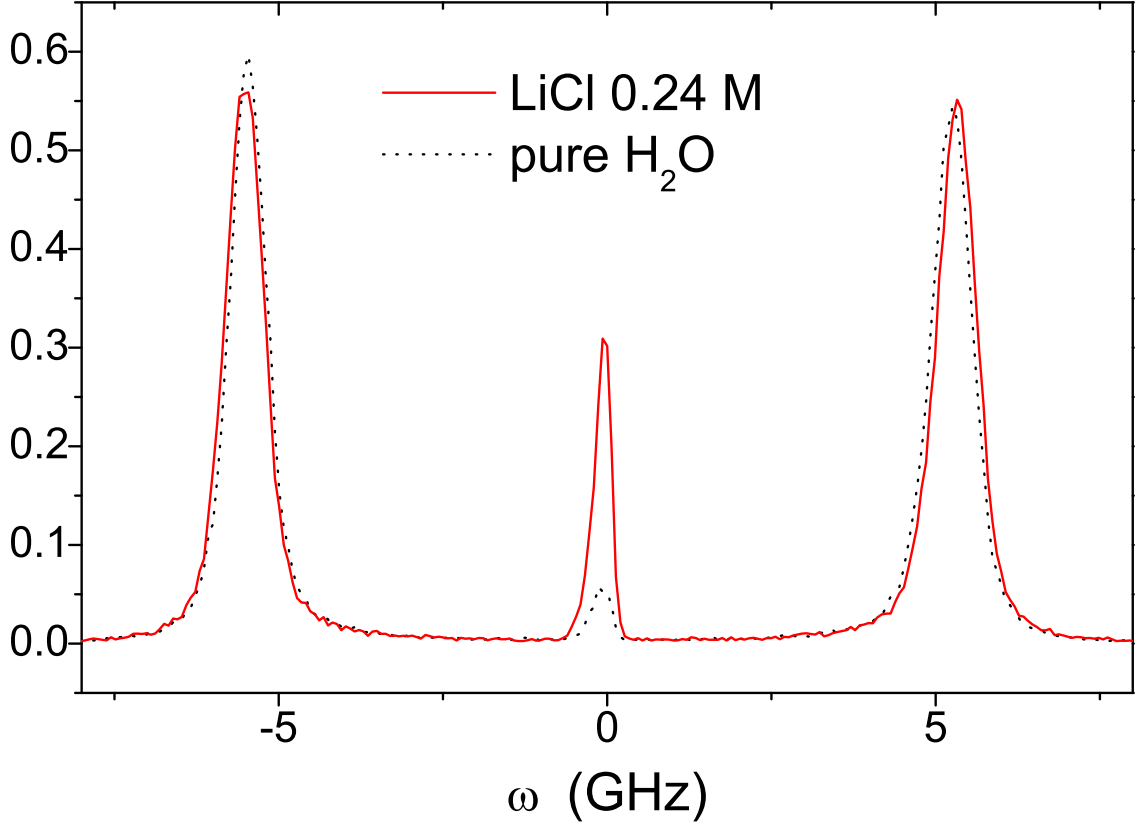}
\caption{\label{fig1} (color online) Rayleigh-Brillouin spectra. Dotted line: pure water; continuous line: C1 LiCl solution, 0.24M.}
\end{figure}

Figure~\ref{fig1} compares the polarized Rayleigh-Brillouin spectrum of pure water to that of the C1 LiCl solution. 

As it will be stated in the discussion, the intense Rayleigh line for the 0.24~M solution is due to the fluctuations of ion concentration. From the Brillouin splitting $2\Delta \nu$ between Stokes and anti-Stokes Brillouin lines, the sound velocity can be determined using Eq.~\ref{eq.3}. With the measured refractive index $n$=1.3328, $\lambda_{\circ}$=532~nm and $\theta$=$\pi/2$, it is found $v$=1475~m/s for pure water, a value that is very close to that obtained earlier \cite{teixeira1}. For the C1 LiCl solution, with the measured $n$=1.3348, one finds $v$=1490~m/s. The increase of sound velocity with respect to that of pure water is $1\%$. The increase of both refractive index and sound velocity are expected because of the increase of respectively the density and of the adiabatic compressibility after the dissolution of of LiCl salt to obtain the 0.24 M solution\cite{uedaira}. It was noted that the $1\%$ variation of the sound velocity is not dependent on the shaking. The approximately same results were obtained with the C1 NaCl solution. From Eq.~\ref{eq.4}, the expected relative variation of the Brillouin intensity $I_{B}$,  caused by the increases of the refractive index and of the sound velocity is of $2\%$. This value lies within the experimental uncertainty for the determination of the $R/B$ ratios. On the other hand, for the highly diluted solutions, from C2 up to C9, no variation of refractive index neither of sound velocity were observed in comparison with those of pure water. From Eq.~\ref{eq.5}, these results allow to conclude that the experimental variation of $R/B$ is a measure of the variation of $I_{R}$. 

\begin{figure}
\includegraphics[width=0.7\textwidth,height=0.7\textheight]{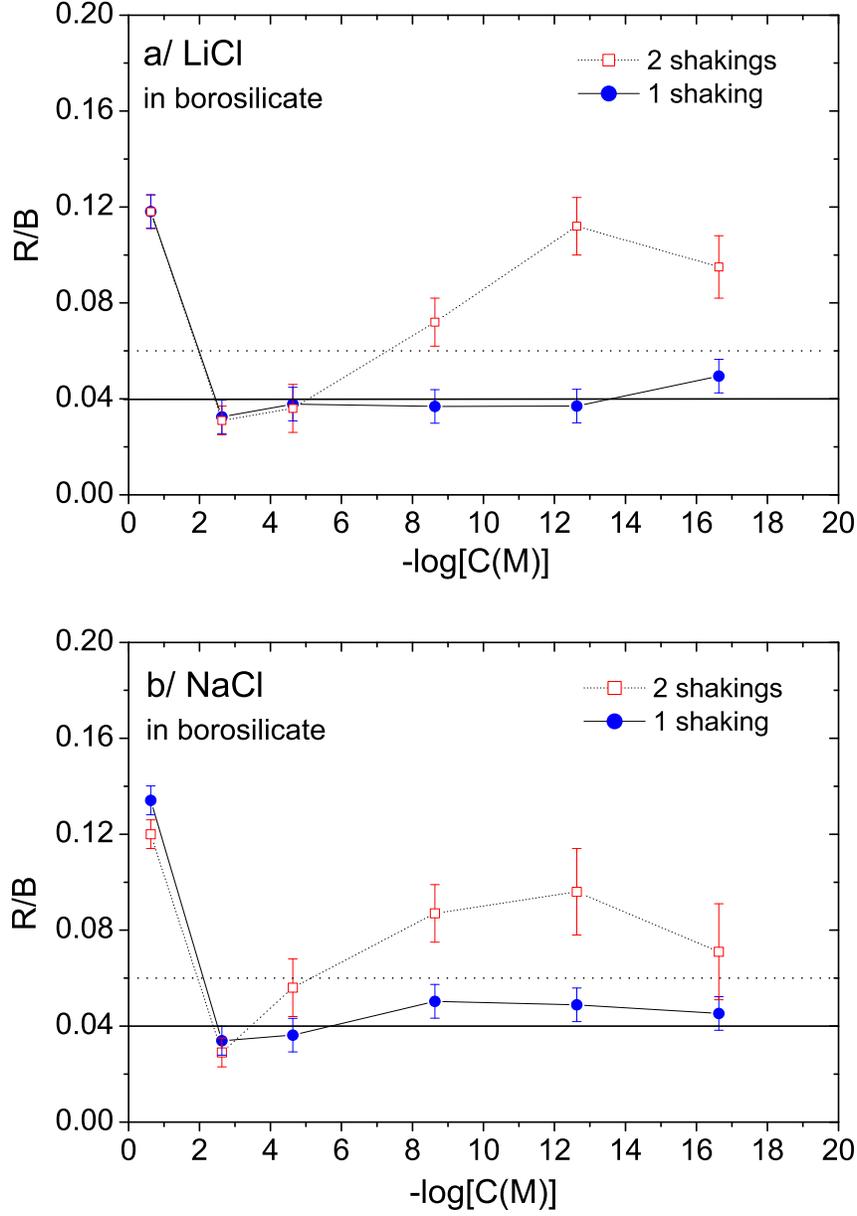}
\caption{\label{fig2} (color online) Landau-Placzek ratio, R/B, versus the inverse of alkali halide concentration C for solutions that were shaken and stored in borosilicate flasks. Full circles: after the dilution-shaking process; empty squares: after the second shaking. \textbf{(a)} LiCl, \textbf{(b)} NaCl. Horizontal continuous line: averaged R/B for reference samples (pure water that was submitted to the same dilution-shaking process as for the solutions); horizontal dotted line: reference samples (after the same second shaking as for the solutions). The error-bars were estimated from several Rayleigh/Brillouin spectra recorded in the space-time of two months after the preparation of the dilutions.}
\end{figure} 

Figure~\ref{fig2} compares the evolutions of the $R/B$ ratios as a function of concentration $C$, as measured one month after the single dilution-shaking process, with those measured from the same solutions which were additionally submitted to a second shaking 5 days after the first dilution-shaking process (the solutions were injected in the quartz cuvettes just after the dilution-shaking process or just after the second shaking, respectively).  It was remarked that the $R/B$ ratios did not change appreciably in the time elapsed between one week and two months after the injection in the quartz cuvettes. The $R/B$ ratios of the LiCl and of the NaCl solutions are compared to the averaged $R/B$ ratios of the reference samples. At this point, it must be noted that the $R/B$ ratios for the non-shaken C1 solutions of LiCl or of NaCl were identical to those of the shaken C1 solutions: this is a proof that in the C1 solutions, the Rayleigh scattering essentially arises from ion concentration fluctuations ($I_{R}(C)$). On the other hand, the observed decrease of the $R/B$ ratios by filtering with filters of different porosities testifies that the Rayleigh (or Rayleigh-Gans) light scattering from diluted solutions with a salt concentration less than 2.10$^{-5}$M is mainly due to submicroscopic heterogeneities ($I_{R}(V)$), whose sizes are approximately distributed between 0.2~$\mu$m and 1~$\mu$m.

\begin{figure}
\includegraphics[width=0.7\textwidth,height=0.7\textheight]{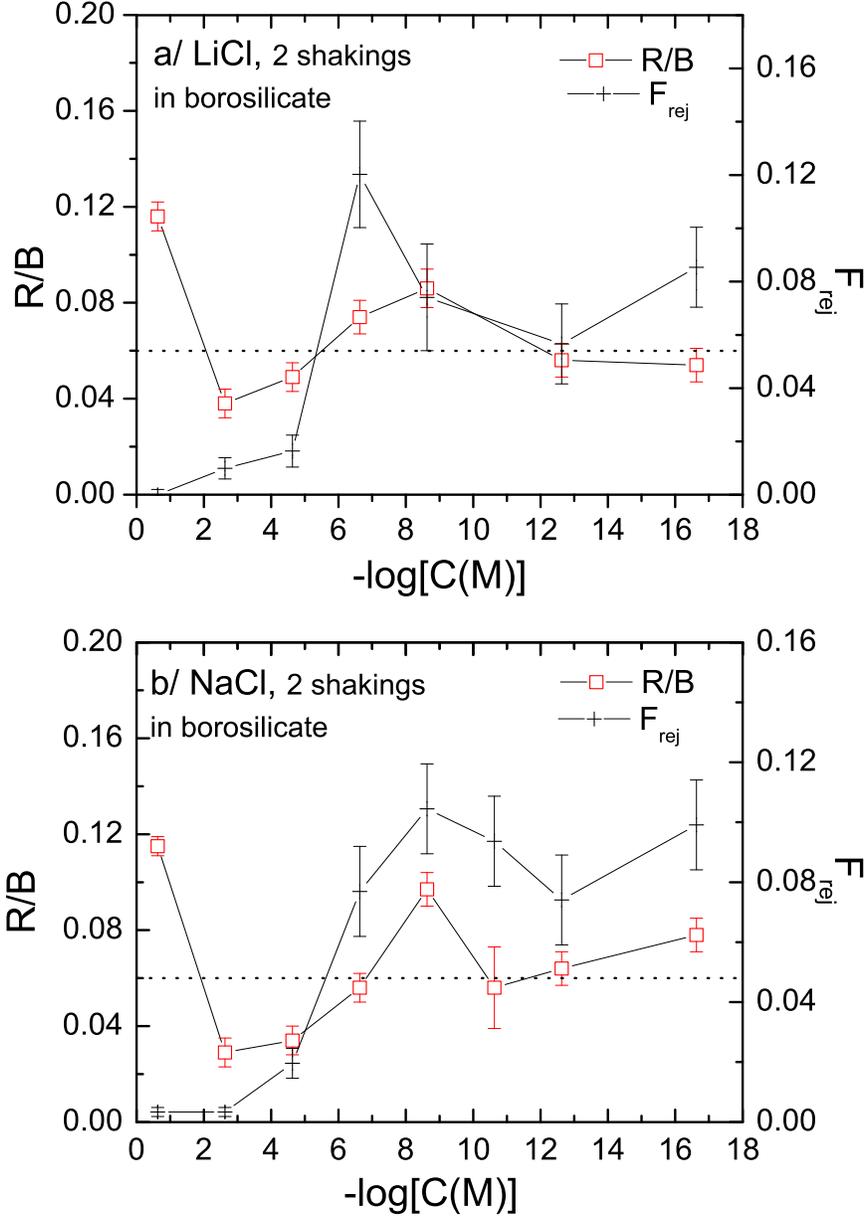}
\caption{\label{fig3} (color online) Landau-Placzek ratio, R/B (empty squares), versus the inverse of alkali halide concentration for another series of dilutions which were submitted to a second shaking 15 days after the first dilution-shaking process. The ratio $F_{rej}$ of rejected spectra (crosses) is also plotted versus the inverse of alkali halide concentration.}
\end{figure}

The main features of the evolution of the $R/B$ ratios as a function of $C$ displayed in Figure~\ref{fig2}  were confirmed for other series of prepared solutions. In Figure~\ref{fig3}, the $R/B$ ratios measured 20 days after the second shaking for another series of diluted solutions, are plotted for LiCl and NaCl solutions that were submitted to a second shaking after the dilution-shaking process, at a later time than for the previous series, i.e. 12 days instead of 5 days. The evolutions of the $R/B$ ratios as a function of $C$ in  Figure~\ref{fig3} are compared to those of the $F_{rej}$ ratio which quantifies the relative number of rejected scans due to flashes occurring from scattering by micrometric objects. It is remarked that $F_{rej}$ is almost zero for concentrations of salt larger than C4 (i.e. $-\log C<4$), and its variation approximately follows that of the $R/B$ ratio for lower concentrations (i.e. $-\log C>4$). The same remarks about $F_{rej}$ were done for the previous series of solutions that are considered in Figure~\ref{fig2}. This observation will be used to help clarify the origin of the heterogeneities (see section IVa).

\begin{figure}
\includegraphics[width=0.7\textwidth,height=0.7\textheight]{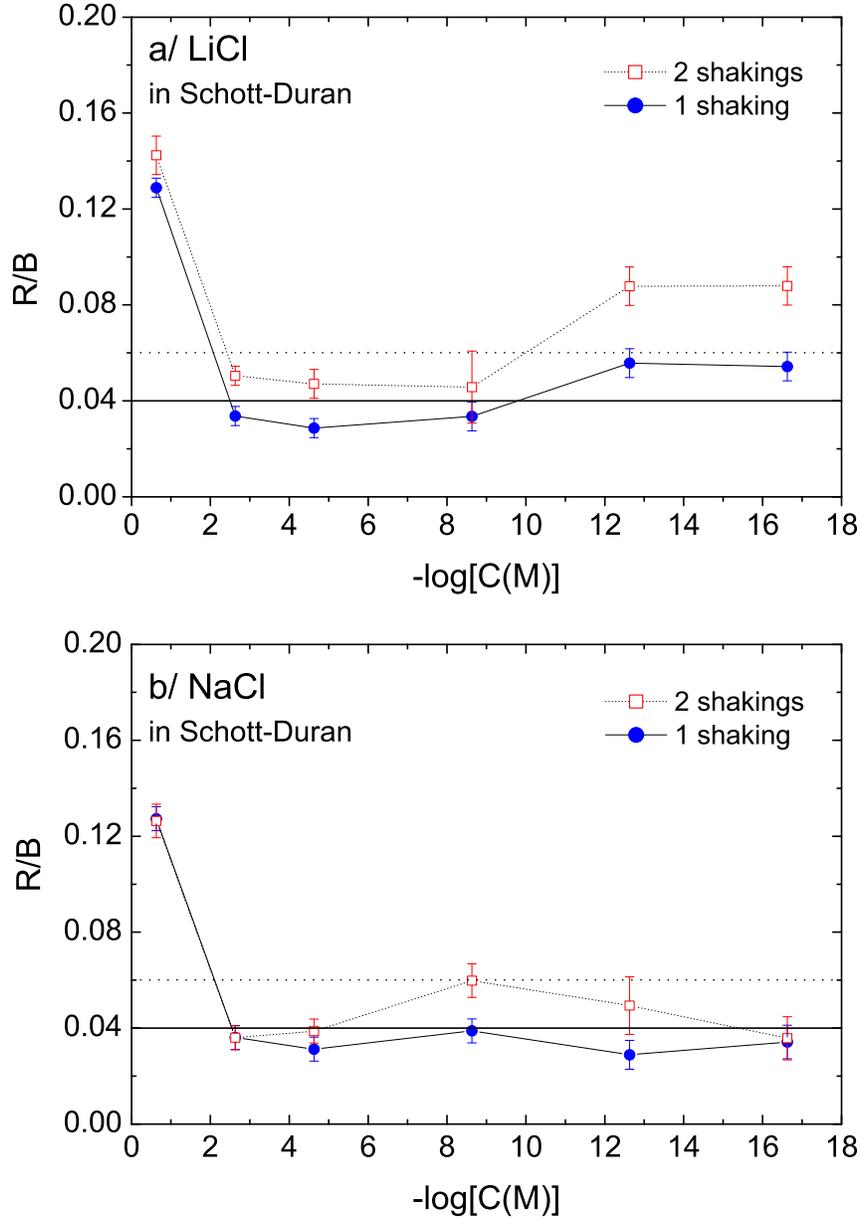}
\caption{\label{fig4} (color online) Landau-Placzek ratio, R/B, versus the inverse of alkali halide concentration C for solutions that were shaken and stored in Schott-Duran flasks. The symbols are the same as in Figure~\ref{fig2}.}
\end{figure}

The $R/B$ ratios shown in Figure~\ref{fig2} and Figure~\ref{fig3} correspond to solutions that were prepared and shaken in borosilicate flasks. In Figure~\ref{fig4}, they are compared to those of solutions that were prepared and shaken in Schott-Duran flasks. For these latter solutions, the light scattering was measured, respectively, 3 days after the first dilution-shaking process and 7 days after the second shaking. In comparison with Figure~\ref{fig2} and Figure~\ref{fig3}, Figure~\ref{fig4} shows that the chemical nature of the glass container has an impact on the observed R/B ratios.   

From Figure~\ref{fig2} to Figure~\ref{fig4}, interesting observations can be pointed out. (1) In addition to the large values of the $R/B$ ratios for the C1 concentrations, that are due to the ion concentration fluctuations, the variations of the $R/B$ ratio as a function of decreasing concentration shows a minimum in the interval C2-C3, followed by a maximum between C5 and C9 depending on the sample. (2) the $R/B$ ratio is increased by the second shaking for the solutions contained in the borosilicate glass flasks while this increase is very much reduced for those contained in Schott-Duran flasks (Figure~\ref{fig4}). The $R/B$ increase observed after the second dilution process was also observed for the pure water reference samples, contained in the borosilicate glasses. This means that the ions released by the walls of borosilicate flasks have an effect on the light scattering. 

\subsubsection{Effect of aging}

\begin{figure}
\includegraphics[width=\columnwidth]{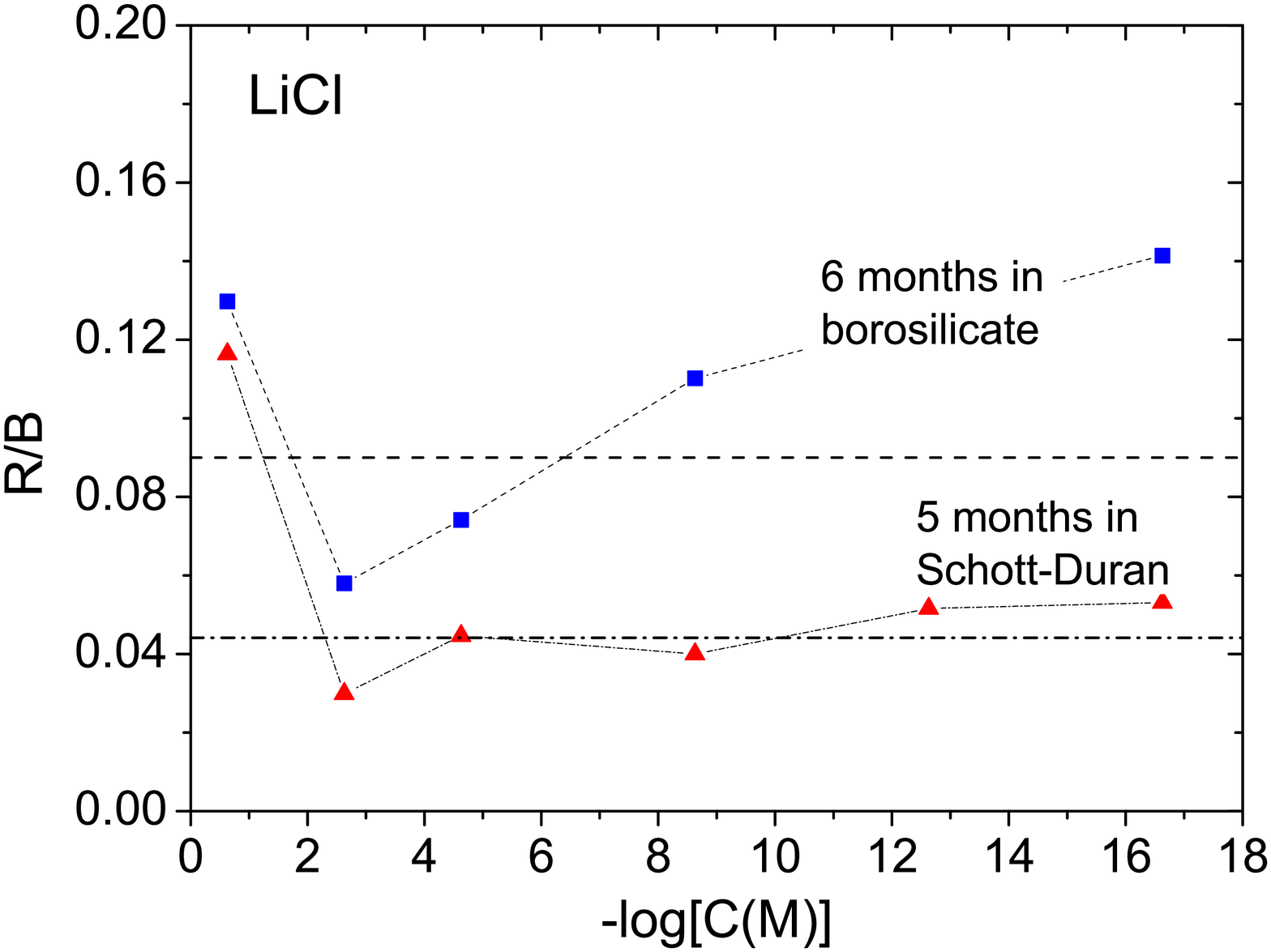}
\caption{\label{fig5} (color online) Landau-Placzek ratio, R/B, versus the inverse of LiCl concentration C for solutions which were submitted to a second shaking, and (1) aged 6 months in borosilicate flasks (squares), (2) aged 5 months in Schott-Duran flasks (triangles). Horizontal dashed line: averaged R/B for pure water that was submitted to the same treatment and the same aging in borosilicate flasks as the LiCl solutions. Horizontal dashed-dotted: averaged R/B for pure water that was submitted to the same treatment and tghe same aging in Schott-Duran flasks as the LiCl solutions. The error-bars are similar to those in Figure~\ref{fig4}.}
\end{figure}

The effect of aging on the $R/B$ ratio depends on the container. Figure~\ref{fig5} compares the $R/B$ ratios of  LiCl solutions kept 6 months at 20$^{\circ}$C (after the second shaking) in borosilicate flasks to those of LiCl solutions kept 6 months at 20$^{\circ}$C but in Schott-Duran flasks; both are compared to the $R/B$ ratios of the reference samples that were shaked and stored in the same condtions. Several informations are obtained from this figure. (1) The average $R/B$ level of pure water is increased after aging in borosilicate flasks (0.06 \textit{vs} 0.09 respectively, comparing Figure~\ref{fig2} with Figure~\ref{fig5}) while it is decreased after aging in Schott-Duran flasks (0.06 \textit{vs} 0.04 respectively, comparing Figure~\ref{fig4} with Figure~\ref{fig5}). (2) The $R/B$ ratios of solutions aged in borosilicate flasks increase after aging, approximately like the reference samples, without any appreciable change of their characteristic evolution with the dilution. (3) The $R/B$ ratios of the solutions aged in Schott-Duran flasks decrease after aging similarly to the reference samples, but the increase at large dilution rates is suppressed. Similar effects of aging were observed with the NaCl solutions.

From the observed effects of aging, it is concluded on the one hand that the ions released from the borosilicate glass during aging increase the light scattering ($I_{R}(V)$) by heterogeneities, both in the salt solutions and in the reference samples. On the other hand, they preserve the submicroscopic heterogeneities created by the salt dissolution and the shaking. The opposite trends observed with the Schott-Duran containers, that release very few ions in the solutions, confirm that the effect observed through Rayleigh-Brillouin scattering intrinsically depend on the ionic content.

\begin{figure}
\includegraphics[width=\columnwidth]{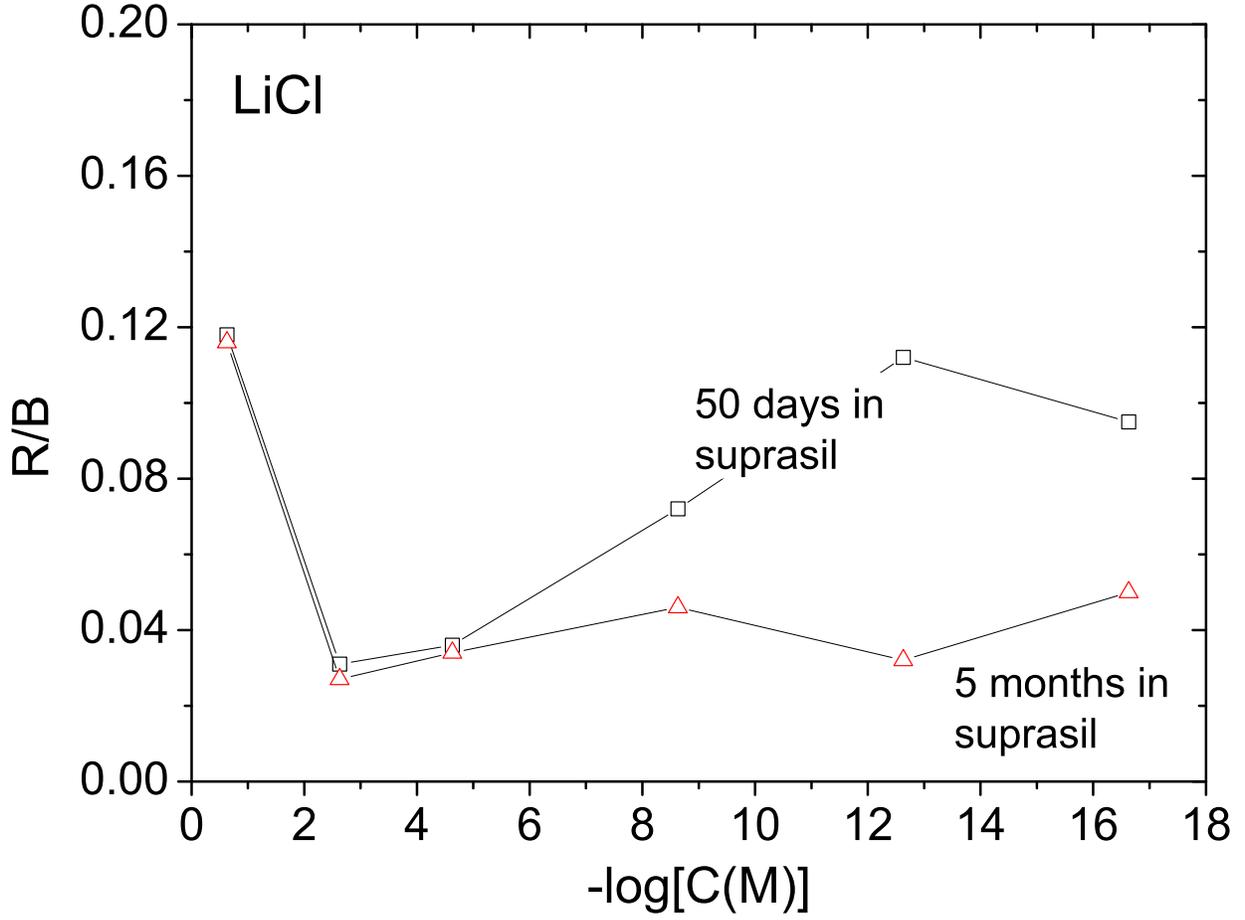}
\caption{\label{fig6} (color online) Landau-Placzek ratio, R/B, versus the inverse of LiCl concentration C for solutions which were submitted to a second shaking, and  aged 5 months in suprasil cuvettes (triangles). It is compared to the R/B of the same solutions, but stored only 50 days in suprasil cuvettes like in Figure~\ref{fig2} (squares). The error-bars are similar to those in Figure~\ref{fig4}.}
\end{figure}

As a confirmation of the aging effect in relation to the container, Figure~\ref{fig6} shows the evolution of the $R/B$ ratios of LiCl solutions aged in Suprasil cuvettes, whose pure silica walls are expected to release small amounts of ions. Like with the Schott-Duran containers, the increase of the $R/B$ ratio at large dilutions rates ($C<2.10^{-5}$M) is found to strongly decrease.

\section{Interpretation}

\subsection{Origin of the Landau-Placzek ratio in shaken aqueous solutions}

\subsubsection{Fluctuations of ion concentrations for $C \geq 2.10^{-5}$M}

The most outstanding experimental result is the variation of the Landau-Placzek ratio, $R/B$, with the ionic concentration $C$, in both LiCl and NaCl solutions. The high level of the $R/B$ ratio for the C1 concentration (0.2M) is clearly related  to $I_{R}(C)$, i.e. the Rayleigh scattering by fluctuations of concentrations of Li$^{+}$, or  Na$^{+}$, and  Cl$^{-}$. Upon decreasing the concentration downto C3 (2.10$^{-5}$M), the $R/B$ ratio decreases, as expected because $I_{R}(C)$ is proportional to the ion concentration  \cite{dubois}. This interpretation is confirmed by the fact that the $R/B$ ratio for highly concentrated solutions is unchanged whether the solutions are shaken or not. Furthemore, it is observed that the $R/B$ ratio for LiCl concentrations larger than 2.10$^{-5}$M is unchanged after aging (Figure~\ref{fig6}). This means that at these high concentrations, the light scattering is largely dominated by the fluctuations of ion concentration, in comparison with a much weaker effect associated with the possible presence of nanobubbles created by the shaking, which becomes detectable at higher dilution rates.

\subsubsection{Submicroscopic bubbles for $C \leq 2.10^{-7}$M}

Surprisingly, the $R/B$ ratio increases when the ion concentration decreases, to reach a maximum in the C5-C9 (2.10$^{-9}$M-2.10$^{-17}$M) concentration range (Figs~\ref{fig2},~\ref{fig3},~\ref{fig5}, ~\ref{fig6}). It must be noted that the minimum of the $R/B$ ratio after the second shaking is always lower than that observed for the reference samples, and the maximum is always higher (Figure~\ref{fig2}). No such variation was observed for non-shaken solutions: the $R/B$ ratio of the highly diluted solutions ($C\leq 2.10^{-7}$M) was approximately the same as that of pure water. It is, therefore, clear that the shaking of the solutions has a decisive effect on the $R/B$ ratio for the highly diluted solutions. Given the extreme care taken in this study about the purity of the samples, it is natural to associate this effect to bubbles created by the vigorous shakings of the solutions upon their preparation.

All the observations described in the previous section prove that the intensity of Rayleigh scattering in the highly diluted solutions is dominated by the intensity $I_{R}(V)$ that is  mainly due to the nanobubles or submicrobubbles which are created by the shakings. 
(1) The possibility that the heterogeneities are clusters of hydrogen-bonded water molecules can be rejected because by using different experimental techniques like low-frequency Raman scattering, or small angle X ray scattering \cite{teixeira2}, one failed to observe any defined clusters in water. Moreover, since $I_{R}(V)$ is proportional to $\Delta n^{2}$, i.e., the square of the difference  between the refractive indices of heterogeneity and of water (Eq.~\ref{eq.2}), it is evident that hypothetical clusters of hydrogen-bonded molecules would have a negligible effect on the $R/B$ ratio in comparison with air bubbles. 
(2) We observed that the ratio $F_{rej}$ (of scans rejected due to scattering flashes) rapidly decreases during the first few days after the shaking. When it is stabilized, its variation with the concentration qualitatively follows  that of the $R/B$ ratio, except in the interval between C1 and C3 where the number of rejected spectra is close to zero (Figure~\ref{fig3}). The objects, which are at the origin of the flashes, and therefore of the rejected spectra, have a micrometric size larger than 2~$\mu$m as indicated in the previous section. The only objects, visible by eye observation through an optical microscope in the highly diluted solutions, were a few moving bubbles with sizes in the range between 2 and 10~$\mu$m. It is then very likely that the origin of the rejected spectra be these micrometric bubbles, whereas the sub-micrometric heterogeneities responsible for the light scattering at lower concentrations are \textit{sub}-microscopic bubbles. In other words, when strong scattering is visible to the eye as flashes, the cause is microscopic bubbles while when strong scattering (i.e. large $R/B$ ratios) is invisible to the eye, it arises from sub-microscopic bubbles, both causes being filiated. It is interesting to remark that $F_{rej}$ is close to zero in the C1-C3 (0.2M-2.10$^{-5}$M) interval, so that in these stabilized solutions at these concentrations, no sub-micrometric bubbles are expected to exist or the bubbles are too small (nanobubbles) to efficiently scatter the light .	
(3) In Figure~\ref{fig6}, one observes that the $R/B$ ratios of solutions kept 5 months in suprasil cuvettes do not change in the C1-C3 range, while they are strongly decreased for $C<2.10^{-7}$~M.

From the values of the R/B ratios observed from solutions filtered with different porosities one may deduce that the diameter $2R$ of the bubbles that are at the origin of the
measured R/B ratios, is 0.2 $\mu$m $\leq 2R\leq$ 1 $\mu$m for $C\leq2.10^{-7}$~M. For $C\geq2.10^{-7}$~M the diameter of the nanobubbles, if they exist, should be $2R\leq 0.2 \mu$m.

\subsection{Stability of the bubbles: role of the ions}.

Before searching the origin of the stability of nanobubbles or submicroscopic bubbles detected by Rayleigh-Brillouin scattering in  diluted solutions of LiCl or NaCl in water, it is emphasised that the absence of observed light scattering by heterogeneities for salt concentrations higher than  2.10$^{-7}$M does not mean that for these concentrations bubbles are absent. From Eq.~\ref{eq.2}, the Rayleigh scattering from heterogeneities is proportional to square of the heterogeneity volume. Therefore, it is possible that nanobubbles, for which the Rayleigh light scattering is negligible because of their too small size,  exist at these concentrations.

The above deduction from experiment, that the air bubbles are responsible for the Rayleigh scattering by heterogeneities in diluted and shaken LiCl or NaCl solutions, poses a problem: because of the surface tension of water, the pressure inside a bubble is inversely proportional to the bubble radius and, in consequence, should be much too high for the existence of long-lived nanometric or sub-micrometric bubbles. This problem can not be resolved by considering the changes of the macroscopic properties of water by  dissolution of salt, because, as it was experimentally demonstrated \cite{weissenborn,craig}, salts such NaCl and LiCl that were dissolved at concentrations less than 10$^{-2}$M, have no effect on the properties of water, like surface tension, viscosity or gas solubilities. As $I_{R}(V)$ depends strongly on the dissolved ions, it is thought that these ions, by their distribution in the solution, have a direct effect on the bubble stability and that the change of their concentration must explain the change of bubble volume.

\subsubsection{Electrically charged bubbles and increase of their volume with the dilution}

The effect of ions on the stability of bubbles in water has been extensively studied by Bunkin and co-workers \cite{bunkin,bunkin97,bunkin2009}. Ions with the  same electric charge, covering the surface of the bubbles, create, through their mutual repulsion, a negative pressure that is opposite to the surface-tension pressure. It was shown that such stabilizing ions at the surface of bubbles are anions \cite{bunkin2009}. The covering of the bubble surface by Cl$^{-}$ anions is consistent with recent molecular dynamics simulations which show that chloride anions have some propensity for the air/water interface, while sodium cations are repelled from this interface \cite{jungwirth}. It is also consistent with the classification of ions according to their partitioning coefficients between bulk water and air-water interface, as deduced from surface tension \cite{pegram}: according to this classification, the concentration of Li$^{+}$ and of Na$^{+}$ at the air-water interface is zero while that of Cl$^{-}$ is markedly different from zero . 

The pressure inside a bubble of radius $R$, that is due to the water surface tension, $\sigma$, is: 

\begin{equation}
\label{eq.6} 
P_{\sigma}=\frac{2\sigma}{R} 
\end{equation}

If $\gamma_{i}$ is the surface density of monovalent ions around the bubble and $\epsilon$ the dielectric constant of water, a simple calculation of electrostatics gives for the negative electric pressure, $P_{E}$, inside the bubble:

\begin{equation}
\label{eq.7}
P_{E}=\frac{e^{2}\gamma_{i}^{2}}{2\epsilon}
\end{equation}

\noindent where $e$ is the electron charge. Neglecting the electric field due to the other ions in water, especially to the solvated cations which diffuse in the space between electrically charged bubbles, the equilibrium is reached when $P_{\sigma}=P_{E}$. In this case, the radius of bubble versus $\gamma_{i}$ is:
 
\begin{equation}
\label{eq.8}
R=\frac{4\sigma\epsilon}{e^{2}\gamma_{i}^{2}}
\end{equation}

\noindent To give an idea of the surface density of ions located around a nanobubble, one chooses $R$=50~nm that is of  the order of the radius estimated by Bunkin et al. \cite{bunkin2009} for a concentration of anions close to 5x10$^{15}$ ions cm$^{-3}$, that is close to C3. With $\sigma=73.10^{-3}$~N/m and with the relative dielectric constant of water $\epsilon_{r}$=80, it is found $\gamma_{i}$=0,4~nm$^{-2}$. The number of ions at the surface of  nanobubble is then, for $R$=50~nm, 1.25.10$^{4}$.

It is interesting to note that from Eq.~\ref{eq.8}  the lower the surface ion density the larger the bubble radius. In order to estimate the consequence of this relation on the $R/B$ ratio, we can make the following hypotheses, that seem realistic at low concentrations of ions: (1) the total volume $NV$ of bubbles per unit volume is constant; (2) the total number of ions surrounding the bubbles in the unit volume is proportional to the ion or salt concentration $C$. Using Eq.~\ref{eq.8}, it can thus be deduced that $N\propto C^{2}$ and $V\propto C^{-2}$. Because, from Eq.~\ref{eq.2}, $I_{R}(V)\propto NV^{2}$, one obtains $I_{R}(V)\propto C^{-2}$. Therefore, by a dilution of a factor 100 of LiCl or of  NaCl in water, $I_{R}(V)$, i.e. the $R/B$ ratio in the highly diluted regime, should be multiplied by 10$^{4}$. This increase of the intensity of light scattering by nanobubbles or sub-micrometric bubbles with the decrease of the square of salt concentration gives only a trend of the evolution of $R/B$ with the dilution : it allows to qualitatively justify the observed increase of the $R/B$ ratio with the dilution rate in the high dilution regime ($C\leq$2.10$^{-7}$M), as well as it accounts for the observed absence of light scattering by bubbles for $C\geq$2.10$^{-5}$M.

\subsubsection{Long bubble lifetime originating in electric interactions}

After having established that the compensation of $P_{\sigma}$ by $P_{E}$ can lead to an increase of bubble volume with the decrease of ion concentration, the consequences of the bubble electric charge on the lifetime bubble assemblies in a given container have to be elucidated. The interaction between bubbles may have several origins \cite{weissenborn}: electric, mechanical, hydrophobic. In presence of ions, it is likely that the interaction between negatively charged nanobubbles is mainly electric. Due to the co-presence of cations in the solutions, the negatively charge nanobubbles may, instead of mutually repelling, assemble through the attractive interaction with the cations that diffuse in the inter-bubble space. The electrostatic bonding between two bubbles through the cation mediation can be relatively strong. Bunkin et al. \cite{bunkin2009} evidenced the presence in water of micrometric aggregates of electrically charged nanobubbles  linked by intermediate cations. The hypothesis of stable negatively charged nanobubbles linked by intermediate cations in aqueous solutions of $\alpha$-cyclodextrin  was also  proposed by  Jin et al. \cite{jin}. However, from a very recent experimental study \cite {ninham}, the light scattering from aggregates of nanobubbles is negligible in comparison with the light scattering from separate bubbles for NaCl concentration lower than $2.10^{-5}$M. This finding supports that for $C<2.10^{-7}$~M the light is scattered from separate bubbles in our very  diluted solutions.

The long lifetime of nanobubbles and the observed dependence on the container of this lifetime (Figures~\ref{fig5} and~\ref{fig6}) can be due to the negative charge of the nanobubbles. By hydrolysis of silica and the dissociation of the silanol SiOH groups into SiO$^{-}$ and H$^{+}$ ions, the glass walls become negatively charged \cite{iler,behrens}, so that the charged nanobubbles are submitted to an electric interaction \cite{bunkin2009} that stabilizes them.
Such interaction is expected to depend on the chemical nature of the surface walls; from our observations, it is expected to be weaker in Schott-Duran containers (and Suprasil cuvettes) than in borosilicate containers. An  additional possible reason for the long-lifetime of the negatively charged nanobubbles can be the affinity of anions for the air-water interface, so that the air/water interface in the cuvettes (meniscus), like in the bubbles, would also be negatively charged. Together with the cuvette walls, this would add stability to the negatively charges bubbles.

\section{Conclusion}

Microscopic and submicroscopic heterogeneities created by shaking of ultrapure aqueous solutions of LiCl and NaCl with dilution ratios covering sixteen decades were studied through Rayleigh-Brillouin scattering. From salt concentrations lower than 2.10$^{-5}$M, the Rayleigh light scattering was found to increase with the decrease of salt concentration, while for concentrations larger than 2.10$^{-3}$M it was found, as expected, to evolve according to concentration fluctuations. All the experimental observations converge to the identification of the submicrometric heterogeneites with  air nanobubbles that were created by the shaking.

As already proposed by different authors \cite{bunkin,jin}, the covering of the bubble surface by negative ions creates an electric pressure which compensates the strong surface tension pressure inside the bubbles, thus stabilizing them. The stabilization of the bubbles by surrounding Cl$^{-}$ anions is consistent with the increase of the Rayleigh light scattering intensity with the decrease of  salt concentration, from concentrations lower than 2.10$^{-5}$M. This increase can be explained by the fact that the surface density of ions around a bubble is inversely proportional to the square-root of the bubble radius, combined with the fact that the intensity of Rayleigh light scattering by submicroscopic bubbles is proportional to the number of bubbles per unit volume $N$ times the square of the bubble volume, $V^2$. 

The long lifetime of the submicroscopic bubbles, that can reach more than 6 months, is probably related to the fact that these bubbles are electrically charged, thereby electrically interacting with the electrically charged walls of the container. Along these lines, the bubble lifetime is expected to change with the electric charge of the walls, and the electric charge of the walls is expected to change with the nature and the amount of ions released in water by these walls. Therefore, the bubble lifetime depends on the container.

The nature of the ions covering the air-water interface of stable nanobubbles is correlated to the classification of ions according to their partioning coefficient between bulk water and air-water interface. As shown by Pegram and Record \cite{pegram} such a classification is close to the Hofmeister series \cite{hofmeister} whereby salts or ions are ranked according to their effectiveness as protein precipitants. A challenging continuation of our works would thus be to study the stability of nanobubbles as a function of the position of the ions in the Hofmeister series.

\begin{acknowledgements}

The authors warmly thank J. Teixeira (CEA/CNRS, Saclay) for very relevant suggestions and careful critical reading of the manuscript, and also M.-P. Pileni for judicious advices. They are very grateful to B. Jourdain and S. Preunkert (LGGE/CNRS, Grenoble) for providing us with ultrapure water. This research was partially financed by Laboratoire Boiron. 

\end{acknowledgements}


\begin{thebibliography}{9}

\bibitem{enderby}
G. W. Neilson, J. E. Enderby,
{\em Proc. R. Soc. London Ser A}
{\bf 390} (1983) 353.

\bibitem{botti}
A. Botti, F. Bruni, S. Imberti, M. A. Ricci, A. K. Sopper,
{\em J. Chem. Phys.}
{\bf 120} (2004) 10154.

\bibitem{bouazizi}
S. Bouazizi, F. Hammami, S. Nasr, M.-C. Bellissent-Funel,
{\em J. Mol. Struct.}
{\bf 892} (2008) 47.

\bibitem{mancinelli}
R. Mancinelli, A. Botti, F. Bruni, M. A. Ricci, A. K. Soper,
{\em J. Phys. Chem. B}
{\bf 111} (2007) 13570.


\bibitem{weissenborn}
P. K. Weissenborn, P. J. Pugh,
{\em J. Colloid Interface Sci.}
{\bf 184} (1996) 550.

\bibitem{craig}
V. S. J. Craig, B. W. Ninham, R. M. Pashley,
{\em J. Phys. Chem.}
{\bf 97} (1993) 10192.

\bibitem{bunkin97}
N. F. Bunkin, A. V. Lobeyev,
{\em Phys. Lett. A}
{\bf 229} (1997) 327.

\bibitem{bunkin2009}
N. F. Bunkin, N. V. Suyazov, A. V. Shkirin, P. S. Ignatiev, K. V. Indukaev,
{\em J. Chem. Phys.}
{\bf 130} (2009) 13408.

\bibitem{hulst}
H. C. van de Hulst,
{\it Light Scattering by Small Particles}
(Dover, New-York, 1981).

\bibitem{cummins}
H. Z. Cummins, R. W. Gammon,
{\em J. Chem. Phys.}
{\bf 44} (1966) 2785.

\bibitem{dubois}
M. Dubois, P. Berge,
{\em Phys. Rev. Lett.}
{\bf 26} (1971) 121.

\bibitem{uedaira}
H. Uedaira, Y. Suzuki,
{\em Bull. Chem. Soc. Jpn.}
{\bf 52} (1979) 2787.

\bibitem{teixeira1}
J. Teixeira, J. Leblond,
{\em J. Physique Lett.}
{\bf 39} (1978) L-83.

\bibitem{teixeira2}
L. Bosio, J. Teixeira, H. E. Stanley,
{\em Phys. Rev. Lett.}
{\bf 46} (1981) 597.

\bibitem{bunkin}
N. F. Bunkin, F. V. Bunkin,
{\em J.E.T.P.}
{\bf 96} (2003) 730.

\bibitem{jungwirth}
P. Jungwirth, D. J. Tobias,
{\em J. Phys. Chem. B}
{\bf 105} (2001) 10468.

\bibitem{pegram}
L. M. Pegram, M. T. Record, Jr,
{\em J. Phys. Chem. B}
{\bf 111} (2007) 5411.

\bibitem{jin}
F. Jin, J. Li, X. Ye, C. Wu,
{\em J. Phys. Chem. B}
{\bf 111} (2007) 11745.

\bibitem{ninham}
N. F. Bunkin, B. W. Ninham, P. S. Ignatiev, V. A. Kozlov, A. V. Shkirin, A. V. Starosvetskij,
{\em J. Biophotonics}
{\bf 4} (2011) 150.

\bibitem{iler}
R. K. Iler,
{\it The Chemistry of Silica}
(Wiley, New-York, 1979).

\bibitem{behrens}
S. H. Behrens, M. Borkovec,
{\em J. Phys. Chem. B}
{\bf 103} (1999) 2918.

\bibitem{hofmeister}
F. Hofmeister,
{\em Arch. Exp. Pathol. Pharmakol.}
{\bf 24} (1888) 247.

\end{thebibliography}
\end{document}